\documentclass[pra,nofootinbib,twocolumn,superscriptaddress,longbibliography]{revtex4-2}
\usepackage{float,graphicx,multirow, amsmath,color,dsfont}
\usepackage{mathtools}
\usepackage{amsmath,bm}
\usepackage{amssymb,mathrsfs,dsfont}
\usepackage{amssymb,mathbbol}
\usepackage{amsfonts}
\usepackage{mathrsfs}
\usepackage{float}
\usepackage{xcolor,hyperref}
\definecolor{darkblue}{rgb}{0,0.0.1,0.3}
\definecolor{darkred}{rgb}{0.6,0.1,0}
\hypersetup{colorlinks,breaklinks,
linkcolor=darkred,urlcolor=darkblue,
anchorcolor=darkred,citecolor=
darkred,pdfauthor=ChSaAr, pdftitle=ChSaAr.qkd.subtraction}
\usepackage{tikz}
\usepackage{mathbbol}
\usepackage{float}
\usepackage[normalem]{ulem}
\usepackage{ulem}

\newcommand{\ie}{\textit{i}.\textit{e}.}

\begin{document}
\title{No  real  advantage of photon subtraction and
displacement in   continuous variable measurement device
independent quantum key distribution}
\author{Chandan Kumar}
\email{chandan.quantum@gmail.com}
\affiliation{Department of Physical Sciences,
Indian
Institute of Science Education and
Research Mohali, Sector 81 SAS Nagar,
Punjab 140306 India.}
\author{Sarbani Chatterjee}
\email{mp18015@iisermohali.ac.in}
\affiliation{Department of Physical Sciences,
Indian
Institute of Science Education and
Research Mohali, Sector 81 SAS Nagar,
Punjab 140306 India.}
\author{Arvind}
\email{arvind@iisermohali.ac.in}
\affiliation{Department of Physical Sciences,
Indian
Institute of Science Education and
Research Mohali, Sector 81 SAS Nagar,
Punjab 140306 India.}
\begin{abstract}
We critically analyse the role of single photon subtraction
(SPS) and displacement in improving the performance of
continuous variable measurement device independent quantum
key distribution (CV-MDI-QKD).  We consider CV-MDI-QKD with
resource states generated by SPS on a displaced two-mode
squeezed vacuum state.  Optimizing the secret key rate with
state parameters reveals that implementing  SPS yields no
benefits in improving the loss tolerance of CV-MDI-QKD.
Additionally, we find that displacement too is not useful in
improving the performance of CV-MDI-QKD.   While our
result is in contradistinction with the widely held belief in the field regarding the
utility of SPS and displacement in CV-MDI-QKD, it also calls
for a re-examination of the role of non-Gaussian operations 
in increasing the efficiency  of various quantum information
processing protocols. 
\end{abstract}
\maketitle
\noindent\textit{Introduction.--} 
Quantum key distribution (QKD) is a quantum communication
process whereby two distant parties, Alice and Bob,
establish a shared secret key in the presence of Eve, a
hostile third party eavesdropper~\cite{gisin-rep-2002,
scarani-rmp-2009,Pirandola-2019}.  The security of such
protocols is guaranteed by the fundamental laws of quantum
physics~\cite{scarani-rmp-2009}.  Depending upon the quantum
systems involved, QKD protocols are  primarily
classified into   discrete variable (DV)
QKD~\cite{Bennett-84, Ekert-prl-1991, Bennett-1992,
Wang2013} and continuous variable (CV) QKD
protocols~\cite{Ralph-pra-1999, Hillery-pra-2000,
Cerf-pra-2001, Grosshans-prl-2002, Grosshans2003,
Weedbrook-rmp-2012,Laudenbach-review}.  Due to its potential
advantages, CV-QKD has attracted significant interest,
leading to the rapid growth of the field, and  several
CV-QKD techniques have been theoretically studied and
experimentally demonstrated in the last two
decades~\cite{Pirandola-2019}.

Although the theoretical foundations of CV-QKD protocols
have been shown to be unconditionally secure, their actual
implementation may be compromised if the assumption of ideal
devices cannot be fulfilled. This might be exploited by an
eavesdropper to covertly glean information about the secret
key.  This problem was addressed by the device-independent
QKD (DI-QKD) protocol, which does away with all presumptions
about the different devices
involved~\cite{diqkd-2005,diqkd-2007}.  However,  the low
secret key rate and short transmission distance render the
DI-QKD  protocol impracticable. Subsequently,
measurement-device-independent  QKD (MDI-QKD) protocols were
proposed for DV systems,  to remove assumptions on the
detectors~\cite{Braunstein-prl-2012, Lo-prl-2012}.
Thereafter, the concept of MDI was extended to the CV
framework~\cite{Pirandola-np-2015, Li-pra-2014,
Xiang-pra-2014}.

While Gaussian states and operations have been at the heart
of CV quantum information
processing~\cite{Weedbrook-rmp-2012,adesso-2007,adesso-2014},
there has been a shift in attention towards non-Gaussian
states and operations~\cite{Mattia}. This can be largely
owed to the fact that non-Gaussianity is essential for
entanglement distillation~\cite{Jarom,Giedke,Eisert} and CV
quantum computation~\cite{Lloyd,Bartlett}. Non-Gaussian
operations have been shown to improve performance of quantum
teleportation~\cite{tel2000,dellanno-2007,tele-2023},
quantum sensing~\cite{josab-2012,metro22} and
QKD~\cite{middle,ebcvqkd,virtual16}.  For instance, single
photon subtraction (SPS) has been shown to enhance the
tolerance to channel losses in CV-MDI-QKD, thereby extending
the maximum transmission
distance~\cite{Ma-pra-2018,chandan-pra-2019,general}.   More
specifically, Ref.~\cite{Ma-pra-2018}  generated an SPS
two-mode squeezed vacuum (TMSV) state  and showed that
tolerance to channel losses can be enhanced.   Subsequently,
Refs.~\cite{chandan-pra-2019,general} employed SPS coupled
with displacement by implementing SPS on  a two-mode
squeezed coherent (TMSC) state in CV-MDI-QKD and showed that
tolerance to channel losses can be further enhanced.

While Refs.~\cite{chandan-pra-2019,general} analysed the
secret key rate   at a  fixed value of the  variance,
transmissivity and displacement parameters of the  SPS-TMSC
state, the  Ref.~\cite{Ma-pra-2018} considered the variance
to be fixed and  optimized  the secret key rate over
transmissivity.  In the present work, we consider the
optimization of the secret key rate  over all the state
parameters aiming to enhance the performance of the
SPS-based CV-MDI-QKD protocol further and to save on
resources.  

Surprisingly, our study reveals that SPS  and displacement
do  not provide any advantage in the optimal situation.
Further, the numerical results shown in
Ref.~\cite{Ma-pra-2018} corresponding to SPS-TMSV state
based CV-MDI-QKD  are incorrect, invalidating their
conclusion claiming enhanced channel loss tolerance.
Similarly, channel loss tolerance enhancement demonstrated
for SPS-TMSC state based CV-MDI-QKD in
Refs.~\cite{chandan-pra-2019,general} was an artifact of
working at high variances. 

 Our findings have broader implications,
while they directly impacts  studies involving
the utility of SPS and displacement in CV-MDI-QKD
protocols~\cite{virtualmdi,amplifiers,underwater} it also
calls for optimization studies to assess the utility of SPS,
and displacement in other CV-QKD protocols including
entanglement-based CV-QKD protocol~\cite{ebcvqkd},
entanglement in the middle CV-QKD protocol~\cite{middle} and
virtual post selection based CV-QKD
protocols~\cite{virtual16}.

The calculation of the covariance matrix of the SPS-TMSC
state required for the evaluation of the secret key rate in
the Hilbert space approach or the Wigner function
approach~\cite{chandan-pra-2019} is challenging and
cumbersome.   Instead, in this work, we employ  the Wigner
characteristic function
formalism~\cite{olivares-2012,adesso-2014} for convenience
and the derivation is provided in the companion paper. 
\begin{figure*} 
\centering
\includegraphics[scale=1]{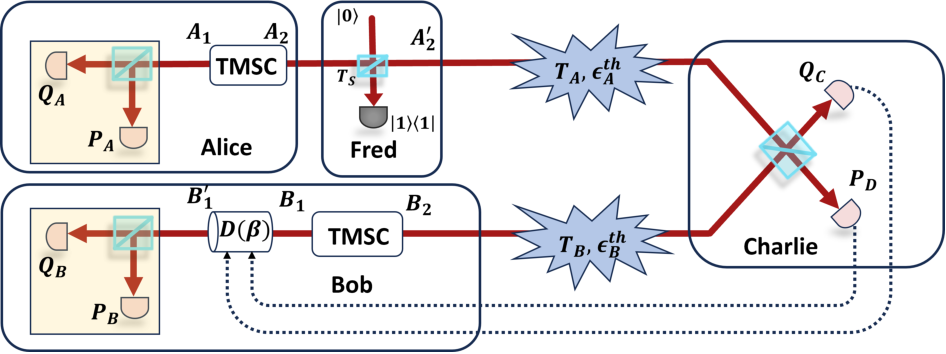} \caption{Schematic
illustration of  SPS-TMSC state based CV-MDI-QKD. The SPS
operation is implemented by Fred using a beam splitter of
transmissivity $T_S$ and a single photon detector.}
\label{mdi_scheme} 
\end{figure*}

\noindent 
\textit{SPS-TMSC state based CV-MDI-QKD.--} 
We recapitulate the SPS-TMSC state based CV-MDI-QKD protocol
here \cite{Li-pra-2014,chandan-pra-2019}.  While the
schematic for the protocol is shown
in Fig.~\ref{mdi_scheme}, the steps involved in the process
of generating a secret key between Alice and Bob under this
protocol are outlined below:

\noindent
{\bf Step 1.}   
Alice prepares the required TMSC state   by the action of
the  two-mode squeezing operator
$S_{12}(r)=\text{exp}[r(\hat{a}^\dagger_{A_1}
\hat{a}^\dagger_{A_2}-\hat{a}_{A_1}\hat{a}_{A_2})]$ on the
coherent state: \begin{equation}\label{tmsc}
|\psi\rangle_{A_1A_2} = S_{12}(r)D_1(d)D_2(d)|00\rangle,
\end{equation} 
where both the modes are  displaced along their respective
$\hat{q}$-quadrature by an amount $d$, \ie, $D_i(d) =
\text{exp}[d ( \hat{a}^\dagger_{A_i} - \hat{a}_{A_i} ) ]$.
The squeezing parameter $r$ is related to the variance $V$
  of the quadrature operators via the relation $V= \cosh
(2r)$~\cite{Weedbrook-rmp-2012,Laudenbach-review}. We obtain
the TMSV state by setting $d=0$ in Eq.~(\ref{tmsc}).

\noindent
{\bf Step 2.} Alice transfers the mode $A_2$ to Fred for
 the  implementation of the SPS operation. In order to facilitate
entanglement swapping, the photon-subtracted mode $A'_2$ is
transferred to Charlie over a quantum channel of length
$L_{AC}$.

\noindent
{\bf Step 3.} Bob prepares a TMSC state of modes $B_1$ and
$B_2$,   similar to the
one prepared by Alice  (with an associated
variance $V$ and displacement $d$)    and  transfers
the latter mode to Charlie through a quantum channel
of length $L_{BC}$.

\noindent
{\bf Step 4:} Using a beam splitter, Charlie mixes the
received modes $A'_2$ and $B_2$, and subsequently subjects
the output modes, labelled as $C$ and $D$, to homodyne
measurements of $\hat{q}$ and $\hat{p}$ operators,
respectively. The  measurement  outcomes
 are denoted as $\lbrace Q_C, P_D\rbrace$, respectively, 
 and are declared by Charlie.

\noindent
{\bf Step 5:}
On the basis of these declared outcomes  by Charlie, Bob
subjects his retained mode $B_1$, to an appropriate
displacement operation $\hat{D}(g(Q_C+ iP_D))$, with $g$
being a gain factor.  Subsequently, the label of the
retained mode is changed to $B'_1$, and marks the completion
of the entanglement swapping process, with the final
scenario of the modes $A_1$ and $B'_1$ being entangled.
Alice and Bob  carry out heterodyne measurements  on their
retained modes to obtain the correlated outcomes, $\lbrace
Q_A, P_A\rbrace$ and $\lbrace Q_B, P_B\rbrace$ respectively.

\noindent
{\bf Step 6:} In the final step classical data post-processing
including information reconciliation (reverse reconciliation)
and privacy amplification  are carried
out to obtain the secret key. 

\noindent
\textit{Secret key rate.--} 
We model the channel  between
Alice and Charlie as a noisy channel
characterized by transmissivity $T_A$ and excess noise
$\varepsilon_A^{\text{th}}$, while the
one between Bob and
Charlie  is characterized by transmissivity $T_B$ and
excess noise $\varepsilon_B^{\text{th}}$. 

 The relation between  the transmissivity $T_A$ and
transmission distance $L_{AC}$ of the quantum channel
between Alice and Charlie is related by   $T_A=10^{- w
L_{AC} /10}$, where  $w$ is the attenuation factor in
decibels per kilometer (dB/km)~\cite{Grasselli2021}. In this
work, we consider $w=0.16$ dB/km~\cite{PRL-2020}. Further,
we approximate the excess noise using a linear fitting as
$\varepsilon_A^{\text{th}}=  19.09 \times 10^{-5}+  6.
13\times 10^{-5} \times L_{AC}$ based on the experimental
values of Ref.~\cite{PRL-2020}.  
Since the CV-MDI-QKD protocol involves two quantum
channels, Eve can implement a correlated two-mode coherent
Gaussian attack, where quantum correlations are injected in
both quantum channels, known as the two-mode
attack~\cite{ottaviani-pra-2015,Pirandola-np-2015}.
However, in a practical scenario, if the two quantum
channels originate from distinct directions, the correlation
between these two channels should  be less. Furthermore,
generating quantum correlations into both channels poses
technical challenges for Eve. Hence, we confine  our study
to two Markovian memoryless Gaussian quantum channels that
operate independently without mutual interaction.  In such a
scenario, the attack by Eve reduces to a one-mode collective
Gaussian attack~\cite{Xiang-pra-2014,Ma-pra-2018}.

Assuming that all of Bob's operations, except for the
heterodyne detection, are not trustworthy, the
  CV-MDI-QKD protocol  described above 
transforms into a  one-way QKD protocol that employs
heterodyne detection~\cite{oneway,prl2004,Li-pra-2014}.  We
emphasize that   the  secret key rate for the equivalent
one-way QKD protocol  is less than or equal to the original
protocol. We use this one-way QKD protocol to calculate
 a bound on the secret key rate for convenience in
calculation~\cite{winter}.

Let $T$ and $\varepsilon^{\text{th}}$ denote the
transmissivity and excess noise of the quantum channel in
equivalent  one-way QKD protocol. The transmissivity $T$ can
be written as  $T=\frac{g^2}{2} T_A$   where $g$ is the gain
of the displacement operation performed by
Bob~\cite{Li-pra-2014}. Further, the optimized excess noise
can be expressed as~\cite{Li-pra-2014}
\begin{equation}
\varepsilon^{\text{th}}=\frac{T_B}{T_A}\left(\varepsilon_B^{\text{th}}
-2\right)+ \varepsilon_A^{\text{th}}+\frac{2}{T_A}.
\label{eq:epsilon_thermal} 
\end{equation}

We can write the
total channel-added noise as $\chi_{\text{ch}} =
\frac{1-T}{T}+\varepsilon^{\text{th}}$.  Let   $I_{AB}$
denote the mutual information between Alice and Bob and
$\chi_{BE}$ denotes the Holevo bound between Bob and Eve,
then the secret key rate of the CV-MDI-QKD protocol  is
given by \begin{equation}
\label{skr}
K=P_{\text{SPS}}\left(\beta I_{AB}-\chi_{BE}\right),
\end{equation} where $P_{\text{SPS}}$ is the success
probability of SPS operation and  $\beta$ is the
reconciliation efficiency.  The analytical expression for
$P_{\text{SPS}}$  is provided in
Refs.~\cite{chandan-pra-2019,general}. The details of the
calculations of mutual information and Holevo bound are
provided in Appendix~\ref{app:skr}.   Further,
the computation of the covariance matrix for the 
SPS-TMSC state, essential for determining the secret key 
rate in either the Hilbert space or Wigner function approach~\cite{chandan-pra-2019},
 presents significant challenges and complexities. 
 However, in this study, we opt for the Wigner characteristic function
  formalism for its practicality and ease of use. The
	derivation  of the Wigner characteristic
	function and the covariance matrix in concise form are provided in the companion paper. 
	The total
transmission
distance turns out to be $L=L_{AC}+L_{BC}$ In this article,
we will consider the scenario when Bob and Charlie are
together ($ L_{BC}=0$ \ie, $T_B=1$) as it renders a high
transmission distance.  The advantage of working with
the key rate give in Eq.~(\ref{skr}) is that one can 
span over various parameter regimes beginning from TMSV to
SPS-TMSC, compare and also optimize the secret key rate.  

\begin{figure}[h!] 
\centering
\includegraphics[scale=1]{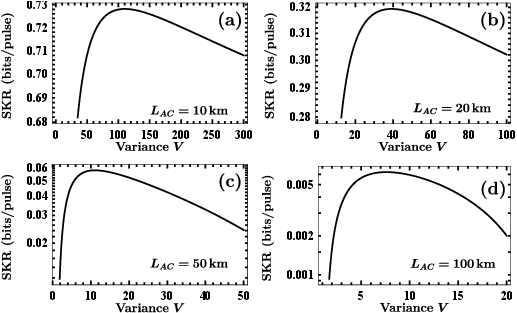}
\caption{Secret key rate (SKR) as a function of variance
for different transmission distance   $L_{AC}$ for TMSV
state based CV-MDI-QKD. We have assumed Charlie and Bob are
at the same place, \ie, $T_B=1$. We have set the numerical
values of the other parameters as $\beta = 96\%$, $T_B=1$,
$\varepsilon_A^{\text{th}}=  19.09 \times 10^{-5}+  6.
13\times 10^{-5} \times L_{AC}  $ and
$\epsilon_B^{\text{th}}=19.09 \times 10^{-5}$. }
\label{tmsv}
\end{figure}

\noindent\textit{Parameter optimization.--} In
Ref.~\cite{Ma-pra-2018}, the secret key was optimized with
respect to transmissivity\footnote{The numerical results of
Ref.~\cite{Ma-pra-2018} are incorrect, which invalidates
their conclusion regarding improved loss tolerance in
CV-MDI-QKD using SPS-TMSV state.}. To motivate the need to
consider the optimization of the secret key rate over
variance, we first analyze the TMSV state based CV-MDI-QKD
and show that variance can be optimized to maximize the
secret key rate. To this end, we plot the secret key rate as
a function of variance at different fixed transmission
distances and show the results in Fig.~\ref{tmsv}. 

We observe that the secret key rate does not increase with
variance; instead, there is an optimal variance that
maximizes the secret key rate. The magnitude of the optimal
variance decreases as we increase the transmission distance. 
This means that increasing the amount of squeezing is not
always helpful and there is an optimal squeezing at which
the protocol displays its best performance.

Strategies such as SPS and displacement was utilized to
enhance the maximum transmission distance or tolerance to
channel loss~\cite{Ma-pra-2018,chandan-pra-2019,general}. 
Here, we revisit these strategies and consider optimization
of state parameters that further enhances the performance.
However, to our surprise, we found that SPS and displacement
operation applied on Alice's side are not advantageous in
enhancing the tolerance to channel loss in CV-MDI-QKD. 

\begin{figure}[h!] 
\centering
\includegraphics[scale=1]{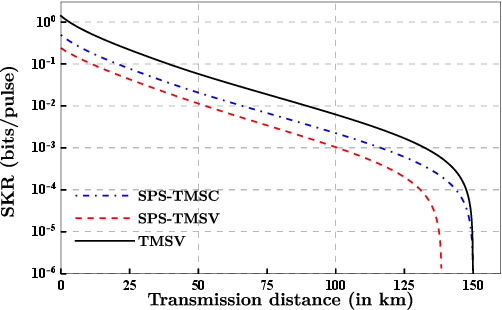}
\caption{  Secret key rate (SKR) as a function of
transmission distance $L_{AC}$. The SKR is optimized over
the state parameters namely variance $V$, displacement $d$
and the transmissivity $T_S$ of the beam splitter involved
in the SPS operation.  We have set the numerical values of
the other parameters   as $\beta = 96\%$, $T_B=1$,
$\varepsilon_A^{\text{th}}=  19.09 \times 10^{-5}+  6.
13\times 10^{-5} \times L_{AC}  $ and
$\epsilon_B^{\text{th}}=19.09 \times 10^{-5}$.}
\label{all}
\end{figure}

The SPS-TMSC state depends on the variance $V$, displacement
$d$, and the beam splitter transmissivity $T_S$ involved in
the implementation of the SPS~\cite{chandan-pra-2019,general}.
Setting  $d=0$ in the SPS-TMSC state, we obtain the SPS-TMSV
state~\cite{Ma-pra-2018}. We optimize the secret key
rate~(\ref{skr}) over these state parameters, which could be
succinctly expressed as
\begin{equation}\label{optimization}
\begin{aligned}
\max_{V,d, T_S} \quad & K(V,d,T_S)\\
\textrm{s.t.} \quad & 1\leq V \leq 15,\\
& 0\leq d \leq 5,\\
&0\leq T_S \leq 1.  \\
\end{aligned}
\end{equation}
Here, we have put an upper limit of $V=15$ on the variance
of the original state\footnote{Squeezed vacuum state with 15
dB squeezing ($V=15.83$) has been demonstrated in
lab~\cite{vahlbruch-prl-2016}.}. We show the optimized
secret key rate as a function of transmission distance
$L_{AC}$ in Fig.~\ref{all}.  

\begin{figure}[h!] 
\centering
\includegraphics[scale=1]{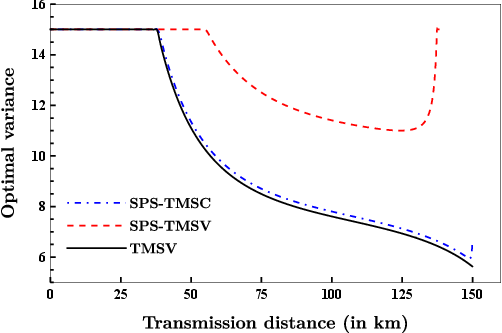}
\caption{ Optimal variance $V$  maximizing the
secret  key rate in Fig.~\ref{all}. }
\label{optimalv}
\end{figure}

We observe the tolerable channel loss for different states
are in the following order:
\begin{equation}
\text{TMSV}  \approx  \text{SPS-TMSC} >  \text{SPS-TMSV}.
\end{equation}
The SPS-TMSV state can tolerate even less channel loss as
compared to the TMSV state.  The SPS-TMSC state, where
displacement is coupled with SPS, can tolerate channel loss
equal to the TMSV state. Therefore,   the SPS operation
implemented on Alice's side provides no advantage in
CV-MDI-QKD at optimal parameters. Further, displacement also
does not help in enhancing the tolerance to channel loss.  
These results are in stark contrast with
Refs.~\cite{Ma-pra-2018,chandan-pra-2019,general}, where SPS and
displacement were claimed to enhance the loss tolerance. 
In Fig.~\ref{optimalv}, we show the variance,
  maximizing the secret  key
rate in the graphs depicted in Fig.~\ref{all}. 

The secret key rate is maximized at variance $V=15$ for
small transmission distances, aligning with the findings in
Fig.~\ref{tmsv} given the optimization constraint  for
variance~(\ref{optimization}). In the limit of maximal
transmission distance, the optimal variance for both TMSV
and SPS-TMSC states converges to $V\approx 6$. These
squeezing variances are easily attainable in laboratory
settings, leading to the conclusion that working with a TMSV
state with a variance of   $V\approx6$ represents the most
favorable choice for CV-MDI-QKD.

 For the SPS-TMSC state, the optimal transmissivity is
slightly below unity and exhibits slight variations. At a
transmission distance of $L_{AC}=25$ km, $T_S=0.995$; at
$L_{AC}=125$ km, $T_S=0.989$. For the SPS-TMSV state,  the
optimal transmissivity   drops well below unity and shows
significant variations as the transmission distance
increases.  Furthermore, the secret key rate is maximized at
a displacement of $d=5$ for almost the entire range of
transmission distance, given the optimization constraint for
displacement~(\ref{optimization}).

\begin{figure}[h!] 
\centering
\includegraphics[scale=1]{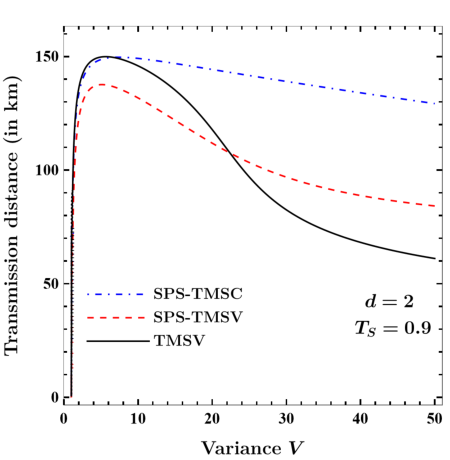}
\caption{  Transmission distance $L_{AC}$ as a function of
variance for a fixed secret key rate $K=10^{-3}$.   We have
set the numerical values of the other parameters as  $\beta
= 96\%$, $T_B=1$,     $\varepsilon_A^{\text{th}}=  19.09
\times 10^{-5}+  6. 13\times 10^{-5} \times L_{AC}  $ and
$\epsilon_B^{\text{th}}=19.09 \times 10^{-5}$.}
\label{contour}
\end{figure}

The advantage shown by SPS-TMSC state in
Refs.~\cite{chandan-pra-2019,general} is an artifact of working at
high variance.  To explicitly show this, we plot the
transmission distance $L_{AC}$ as a function of variance for
a fixed secret key rate $K=10^{-3}$ and show the result in
Fig.~\ref{contour}.  We note that
Refs.~\cite{chandan-pra-2019,general} showed the enhanced loss
tolerance for SPS-TMSC state based CV-MDI-QKD at $V=50$,
$T_S=0.9$, and $d=2$. 

Clearly,  in the high variance regime (for instance,
$V=50$), the maximum transmission distance of the considered
states in CV-MDI-QKD can be arranged in the following order: 
\begin{equation}
\text{TMSV} <  \text{SPS-TMSV}  <   \text{SPS-TMSC},
\end{equation}
which was precisely shown in Refs.~\cite{chandan-pra-2019,general}.
However, we notice that by working at $V\approx 6$, the
performance of different states are in the following order:
\begin{equation}
\text{TMSV}  \approx  \text{SPS-TMSC} >  \text{SPS-TMSV}.
\end{equation}
This result further corroborates with our main result shown
in Fig.~\ref{all}.  Therefore,  analysis performed at
non-optimal regime (high variance) in
Refs.~\cite{chandan-pra-2019,general}  leads to an
appearance of advantage with displacement and SPS. However,
this advantage is merely apparent and not genuine.

\noindent
\textit{Conclusion.--} 
While earlier studies utilizing SPS and displacement
CV-MDI-QKD have been conducted at fixed value of state
parameters, we  optimized the secret key rate over various
state parameters with a view to optimize the performance
and to obtain a global picture.
Surprisingly, we discovered that neither SPS nor
displacement offered any advantage in enhancing the loss
tolerance of CV-MDI-QKD. The observed enhancement in the
loss tolerance demonstrated for CV-MDI-QKD with the SPS-TMSC
state  could be attributed to
operating at high variance, which is
non-optimal~\cite{chandan-pra-2019,general}.  Moreover, the
numerical results presented in Ref.~\cite{Ma-pra-2018}
regarding CV-MDI-QKD with the SPS-TMSV state were found to
be inaccurate, thereby refuting their assertion of enhanced
loss tolerance through SPS.  Our result will also impact
similar studies in
CV-MDI-QKD~\cite{virtualmdi,amplifiers,underwater}.
Therefore, our findings questions a commonly accepted notion
within the field that SPS and displacement are beneficial in
enhancing loss tolerance.

Our work raises numerous questions regarding the utility of
non-Gaussian operations in CV quantum information
processing, which is currently pursued by a large community
of researchers.  While we have focused on SPS by Alice, it
is essential to also consider the scenario of SPS by Bob.
While the utility of SPS in other
CV-QKD protocols~\cite{middle,ebcvqkd,virtual16} needs to be
properly re-assessed,  the
potential benefits of other non-Gaussian operations such as
photon catalysis~\cite{catpra2019,virtualzpccat,general} and
photon addition~\cite{addition,general}  in CV-QKD
also require critical examination.   
Since SPS  has been shown to be useful in quantum
teleportation~\cite{tel2000,dellanno-2007,tele-2023} and
quantum metrology~\cite{josab-2012,metro22} our work call
for a re-examination of these results from an optimality
point of view.

\textit{Acknowledgement.---} 	S.C. acknowledges the Prime
Minister's Research Fellowship (PMRF) scheme, GoI, for
financial support. A. and C.K.   acknowledge
the financial support from {\bf
DST/ICPS/QuST/Theme-1/2019/General} Project number {\sf
Q-68}. 

\appendix

\section{Calculation of the mutual information and Holevo
	bound} \label{app:skr}
We start by computing the mutual information.  We utilize
the covariance matrix of the SPS-TMSC state to calculate the
secret key rate, which is a lower bound of the secret key
rate of the non-Gaussian state~\cite{Cerf-prl-2006}. The
covariance matrix of the SPS-TMSC state can be expressed in
the following form:
\begin{equation}\label{app:cov}
	\Sigma_{A_1 A'_2} =
	\begin{pmatrix}
		V_{A}^{q} & 0 & V_{C}^{q} & 0 \\
		0 & V_{A}^{p} & 0 & V_{C}^{p} \\
		V_{C}^{q} & 0 & V_{B}^{q} & 0 \\
		0 & V_{C}^{p} & 0 & V_{B}^{p}
	\end{pmatrix},
\end{equation}
where $({\Sigma_{A_1 A'_2}})_{ij} =
\frac{1}{2}\langle\{ \hat{\xi_i},\hat{\xi_j} \} \rangle -
\langle \hat{\xi_i}\rangle \langle \hat{\xi_j}\rangle$.
 
The  covariance matrix representing the state of the
modes $A_1 B'_1$ after interaction with the noisy channel
can be written as~\cite{Ma-pra-2018}
\begin{equation}
	\begin{aligned}
		\Sigma_{A_1 B'_1} = &\begin{pmatrix}
			V_{A}^{q} & 0 & \sqrt{T}V_{C}^{q} & 0 \\
			0 & V_{A}^{p} & 0 & \sqrt{T}V_{C}^{p} \\
			\sqrt{T}V_{C}^{q} & 0 & T(V^q_{B}+\chi_\text{ch}) & 0 \\
			0 & \sqrt{T}V_{C}^{p} & 0 & T(V^p_{B}+\chi_\text{ch})
		\end{pmatrix},\\
		&=\begin{pmatrix}
			\delta_1 & 0 & \kappa_1 & 0 \\
			0 &\delta_2 & 0 & \kappa_2 \\
			\kappa_1 & 0 & \mu_1 & 0 \\
			0 & \kappa_2& 0 & \mu_2 \\
		\end{pmatrix}
		=\begin{pmatrix}
			\Sigma_{A_1} & \Sigma_{C} \\
			\Sigma_{C} & \Sigma_{B'_1}\\
		\end{pmatrix}.
		\label{eq:variance_a1b1}
	\end{aligned}
\end{equation}
The covariance matrix of Alice is computed as follows after
Bob performs heterodyne measurement:
\begin{equation}
	\begin{aligned}
		\Sigma_{A_1|B'_1} =
		&\Sigma_{A_1}-\Sigma_C\left(\Sigma_{B'_1}+\mathbb{1}_2\right)^{-1}(\Sigma_C)^T,\\
		=&\begin{pmatrix}
			\delta_1-\frac{\kappa_1^2}{\mu_1+1} & 0 \\
			0 & \delta_2-\frac{\kappa_2^2}{\mu_2+1} \\
		\end{pmatrix}.
	\end{aligned}
	\label{eq:varianceagivenb}
\end{equation}
The classical mutual information is given by
\begin{equation}
	I_{AB}=\frac{1}{2}\log_2\left(\frac{\Sigma_{A_1}^q+1}{\Sigma_{{A_1}|B'_1}^q+1}\right)+
	\frac{1}{2}\log_2\left(\frac{\Sigma_{A_1}^p+1}{\Sigma_{{A_1}|B'_1}^p+1}\right),
	\label{eq:mutualinfo}
\end{equation}
For the computation of the Holevo bound, we make the
assumption that Eve possesses a purification of the state
$\hat{\rho}_{A_1B'_1EF}$ and has access to Fred's mode. The
Holevo bound is then determined as follows:
\begin{equation}
	\begin{aligned}
		\chi_{BE} = S(\hat{\rho}_{A_1 B'_1}) - S(\hat{\rho}_{A_1|B'_1}).
		\label{eq:holevo}
	\end{aligned}
\end{equation}
Here, $S(\hat{\rho})$ represents the von Neumann entropy
associated with the state $\hat{\rho}$, and it is defined as
follows:
\begin{equation}
	S(\hat{\rho}) =  \Sigma_i g(\nu_i),
	\label{eq:rho}
\end{equation}
where,
\begin{equation}
	g(\nu) =
	\frac{\nu+1}{2}\log_2\left(\frac{\nu+1}{2}\right)-
	\frac{\nu-1}{2}\log_2\left(\frac{\nu-1}{2}\right).
	\label{eq:rho_g}
\end{equation}
Here, the symplectic eigenvalues of the covariance matrix of
the state $\hat{\rho}$ are denoted as $\nu_i$. The
calculation of $S(\hat{\rho}_{A_1 B'_1})$ involves these
symplectic eigenvalues obtained from the matrix in
Eq.~(\ref{eq:variance_a1b1}), which are determined as
follows:
\begin{equation}
	\nu_{1,2} = \frac{1}{\sqrt{2} }\left[I_1+I_2+2I_3 \pm
	\sqrt{(I_1+I_2+2I_3)^2-4 I_4}\right]^{1/2},
\end{equation}
where $I_1=\delta_1 \delta_2$, $I_2=\mu_1 \mu_2$,
$I_3=\kappa_1 \kappa_2$, and
$I_4=(\delta_1 \mu_1-\kappa_1^2)(\delta_2 \mu_2-\kappa_2^2)$.
The computation of $S(\hat{\rho}_{A_1|B'_1})$ utilizes the
symplectic eigenvalue extracted from the matrix provided in
Eq.~(\ref{eq:varianceagivenb}), yielding the following
result:
\begin{equation}
	\nu_3 = \sqrt{\left(
		\delta_1-\frac{\kappa_1^2}{\mu_1+1}\right)\left(
		\delta_2-\frac{\kappa_2^2}{\mu_2+1}\right)}.
\end{equation}
The secret key rate~(\ref{skr}) can be calculated using the
expressions of mutual information and Holevo bound.

%

\end{document}